\documentclass[fleqn,12pt,twoside]{article}
\usepackage{espcrc1}
\usepackage{epsfig}
\usepackage{amsmath}

\def\la{\mathrel{\mathpalette\fun <}}

\def\fun#1#2{\lower3.6pt\vbox{\baselineskip0pt\lineskip.9pt
  \ialign{$\mathsurround=0pt#1\hfil##\hfil$\crcr#2\crcr\sim\crcr}}}

\def\tform{t_{\mbox{\scriptsize form}}}

\def\lrang#1{\left\langle#1\right\rangle}
\def\cO#1{{\cal O}\left(#1\right)}

\def\sigmatot{\sigma^{\mbox{\scriptsize tot}}}

\def\1{$^{\mbox{\scriptsize\{1\}}}$}
\def\2{$^{\mbox{\scriptsize\{2\}}}$}

\def\np#1#2#3{{ Nucl.Phys.}~\underline{B#1} (19#3) #2}
\def\pl#1#2#3{{ Phys.Lett.}~\underline{#1B} (19#3) #2}

\def\prl#1#2#3{{ Phys.Rev.Lett.}~\underline{#1} (19#3) #2}

\title{QCD at moderately large distances}

\author{Yuri Dokshitzer\thanks{On leave 
from PNPI, 188350 Gatchina, St.~Petersburg, Russia.} \\ { LPTHE,
Universit{\'e} Pierre et Marie Curie,\\ 4, Place Jussieu, 75252 Paris,
France }}
       
\begin{document}

\maketitle

\begin{abstract}
\end{abstract}

\section{INTRODUCTION}

Phenomenology of scattering processes involving hadrons always was,
and still is, providing puzzles and inspiration.  
If 30--40 years ago quantum field theory (QFT) had been kept in higher
respect, the most general phenomenological features of hadron
interactions that were known then could have already hinted at QCD as
a possible underlying microscopic theory of hadrons.

Hints from the past:
\begin{itemize}
\item 
The fact that in high energy hadron interaction processes {\em
inelastic}\/ breakup typically dominates over elastic scattering
hinted at proton being a loosely bound compound object:
\begin{quote}
$\Longrightarrow$\qquad   {\em Constituent Quarks}
\end{quote}
 
\item
Constancy of transverse momenta of produced hadrons, rare appearance
of large-$k_\perp$ fluctuations, was signaling the weakness of
interaction at small relative distances:
\begin{quote}
$\Longrightarrow$\qquad  {\em Asymptotic Freedom}
\end{quote}

\item 
The last but not the least; 
\begin{itemize}

\item
The total hadron interaction cross sections turned out to be
practically {\em constant}\/ with energy. {\em If}\/ we were to employ
the standard quantum field theory (QFT) picture of a particle exchange
between interacting objects,
$$
   \sigma_{\mbox{\scriptsize tot}} \propto s^{J-1} \>\simeq\> \mbox{const},
$$
{\em then} this called for a spin-one elementary field, $J=1$, to be
present in the theory.

\item
{\em Uniformity in rapidity}\/ of the distribution of produced hadrons
(Feynman plateau) pointed in the same direction, {\em if}, once again,
we were willing to link final particle production to accompanying QFT
``radiation''.
\end{itemize}
\begin{quote}
$\Longrightarrow$\qquad  {\em Vector Gluons}.
\end{quote}
\end{itemize}
Nowadays the dossier of puzzles \&\ hints that the hadron
phenomenology has accumulated is very impressive. It includes a broad
spectrum of issues ranging from unexplained regularities in hadron
spectroscopy to soft ``forceless'' hadroproduction in hard processes.
To locate and formulate a puzzle, to digest a hint, --- these are the
road-signs to the hadron chromodynamics construction site.  We are
learning how to listen. And to hear.

The reason why one keeps talking, for almost 30 years now, about
puzzles and hints, about {\em constructing}\/ QCD rather than {\em
applying}\/ it, lies in the conceptually new problem one faces when
dealing with a non-Abelian theory with unbroken symmetry (like
QCD). We have to understand how to master QFTs whose dynamics is
intrinsically unstable in the infrared domain: the objects belonging
to the physical spectrum of the theory (supposedly, colourless hadrons,
in the QCD context) have no direct one-to-one correspondence with the
fundamental fields the microscopic Lagrangian of the theory is made of
(coloured quarks and gluons).

In these circumstances we don't even know how to formulate at the
level of the microscopic fields the fundamental properties of the
theory, such as conservation of probability (unitarity) and
analyticity (causality):
\begin{itemize}
\item
  What does {\em Unitarity}\/ imply for confined objects?
\item
  How does {\em Causality}\/ restrict quark and gluon Green functions
  and their interaction amplitudes?
\item
  What does the {\em mass}\/ of an INFO (Identified Non-Flying Object)
  mean?
\end{itemize}
The issue of quark masses is especially damaging since a mismatch
between quark and hadron thresholds 
% (whatever the former might mean)
significantly affects predicting the yield of heavy-flavoured hadrons
in hadron collisions. 
% markedly at LHC.

Understanding confinement of colour remains an open problem.  Given
the present state of ignorance, one has no better way but to circle
along the {\em Guess-Calculate-Compare}\/ loop.  There are, however,
{\em guesses}\/ and {\em guesses}.

\section{WORDS, WORDS, ...}

Speaking of ``perturbative QCD'' (pQCD) can have two different
meanings.
\begin{itemize}
\item
 In a narrow, strict sense of the word, {\em perturbative approach}\/
 implies representing an answer for a (calculable) quantity in terms
 of series in a (small) expansion parameter $\alpha_s(Q)$, with $Q$
 the proper hardness scale of the problem under consideration.

\item
 In a broad sense, {\em perturbative}\/ means applying the language of
 quarks and gluons to a problem, be it of perturbative
 (short-distance, small-coupling) or even non-perturbative nature.
\end{itemize}

The former definition is doomed: the perturbative series so
constructed are known to diverge. In QCD these are asymptotic series
of such kind that cannot be ``resummed'' into an analytic function in
a unique way. For a given calculable (collinear-\&\ {}-infrared-safe;
CIS) observable~\cite{SW} the nature of this nasty divergence can be
studied and quantified as an intrinsic uncertainty of pQCD series, in
terms of the so-called {\em infrared renormalons}\/~\cite{Beneke}.
Such uncertainties are non-analytic in the coupling constant and
signal the presence of non-perturbative (large-distance) effects.  For
a CIS observable, non-perturbative physics enters at the level of
power-suppressed corrections $\exp\{-c/\alpha_s(Q)\}\propto
Q^{-\gamma}$, with $\gamma$ an observable-dependent positive
integer\footnote{usually, though not necessarily~\cite{EEC}} number.

On the contrary, the broader definition of being ``perturbative'' is
bound to be right. At least as long as we aim at eventually deriving
the physics of hadrons from the quark-gluon QCD Lagrangian.

To distinguish between the two meanings, in what follows we will
supply the word {\em perturbative}\/ with a superscript \1  or \2.  
Thus, when discussing the strong interaction domain in terms of quarks
and gluons, we will be able to speak about non-perturbative\1
perturbative\2 effects.

\section{PROBING CONFINEMENT WITH PERTURBATIVE TOOLS}
% {Probing confinement with perturbative tools}

Let us discuss the test case of the total cross section of $e^+e^-$
annihilation into hadrons as an example.

To predict $\sigma^{\mbox{\scriptsize tot}}_{\mbox{\scriptsize hadr}}$
one calculates instead the cross sections of quark and gluon
production, $(e^+e^-\to q\bar{q})$ + $(e^+e^-\to q\bar{q}+g)$ + etc.,
where quarks and gluons are being treated {\em perturbatively}\/ as
real (unconfined, flying) objects. The {\em completeness}\/ argument
provides an apology for such a brave substitution:
\begin{quote}
  Once instantaneously produced by the electromagnetic (electroweak)
  current, the quarks (and secondary gluons) have nowhere else to go
  but to convert, {\em with unit probability}, into hadrons in the end
  of the day.
\end{quote}
This {\em guess}\/ looks rather solid and sounds convincing, but
relies on two hidden assumptions:
\begin{enumerate}
\item 
  The allowed hadron states should be numerous as to provide the
  quark-gluon system the means for ``regrouping'', ``blanching'',
  ``fitting'' into hadrons.
\item
  It implies that the ``production'' and ``hadronization'' stages of
  the process can be separated and treated independently.
\end{enumerate}
1. To comply with the first assumption the annihilation energy has to
be taken large enough, $s\equiv Q^2\gg s_0$. In particular, it fails
miserably in the resonance region $Q^2\la s_0\sim 2M_{\rm res}^2$.
Thus, the point-by-point correspondence between hadron and quark cross
sections,
$$
 \sigma^{\mbox{\scriptsize tot}}_{\mbox{\scriptsize hadr}}(Q^2) 
 \>\stackrel{\mbox{?}}{=}\> 
 \sigma^{\mbox{\scriptsize tot}}_{{q\bar{q}+X}}(Q^2), 
$$
cannot be sustained except at very high energies.
%%%%% How high is high? 
%%%%% And how accurate? 
It can be traded, however, for something better manageable.

Invoking the dispersion relation for the photon propagator (causality
$\Longrightarrow$ analyticity) one can relate the {\em energy
integrals}\/ of $\sigmatot(s)$ 
% over the Minkowsky region 
with the correlator of electromagnetic currents in a deeply Euclidean
region of large {\em negative}\/ $Q^2$. The latter
% which 
corresponds to small space-like distances between interaction points,
where the perturbative\2 approach is definitely valid.

Expanding the answer in a formal series of local operators, one
arrives at the structure in which the corrections to the trivial unit
operator generate the usual perturbative\1 series in powers of
$\alpha_s$ (logarithmic corrections), whereas the vacuum expectation
values of dimension-full (Lorentz- and colour-invariant) QCD operators
provide non-perturbative\1 corrections suppressed as powers of $Q$.

This is the realm of the famous ITEP sum rules~\cite{ITEP}
which proved to be successful in linking the parameters of the
low-lying resonances in the Minkowsky space with expectation values
characterising a non-trivial structure of the QCD vacuum in the
Euclidean space.
The leaders among them are the gluon condensate $\alpha_s
G^{\mu\nu}G_{\mu\nu}$ and the quark condensate
$\lrang{\psi\bar\psi}\lrang{\psi\bar\psi}$ which contribute to the
total annihilation cross section, symbolically, as
\begin{equation}
\label{eq:itep}
\sigma^{\mbox{\scriptsize tot}}_{\mbox{\scriptsize hadr}}(Q^2) 
-  \sigma^{\mbox{\scriptsize tot}}_{{q\bar{q}+X}}(Q^2)
\>=\> c_1{\frac{\alpha_s G^2}{Q^4} + c_2\frac{\lrang{\psi\bar\psi}^2}{Q^6}} 
+ \ldots\,.  
\end{equation}

\noindent
2. Validating the second assumption also calls for large $Q^2$.  To be
able to separate the two stages of the process, it is {\em
necessary}\/ to have the production time of the quark pair $\tau\sim
Q^{-1}$ to be much smaller than the time $t_1\sim \mu^{-1}\sim 1\,{\rm
fm}/c$ when the first hadron appears in the system. Whether this
condition is {\em sufficient}, is another valid question. And a tricky
one.

Strictly speaking, due to gluon bremsstrahlung off the primary quarks,
the perturbative production of secondary gluons and $q\bar{q}$ pairs
spans an immense interval of time, ranging from a very short time
$\tform\sim Q^{-1}\ll t_1$ 
all the way up to a macroscopically large time 
$\tform\la Q/\mu^2\gg t_1$.

This accompanying radiation is responsible for formation of hadron
jets. It does not, however, affect the total cross section.  It is the
rare hard gluons with large energies and transverse momenta,
$\omega\sim k_\perp\sim{Q}$, that only matter.  This
% statement constitutes the essence of 
follows from 
the famous Bloch-Nordsieck theorem
%%%%% ~\cite{BN} 
which states that the logarithmically enhanced (divergent)
contributions due to real production of {\em collinear}\/ ($k_\perp\ll
Q$) and {\em soft}\/ ($\omega\ll Q$) quanta cancel against the
corresponding virtual corrections:
$$
 \sigma^{\mbox{\scriptsize tot}}_{q\bar{q}+X} = \sigma_{Born}
\left(1+\frac{\alpha_s}{\pi}\left[\infty_{\mbox{\scriptsize real}}
-\infty_{\mbox{\scriptsize virtual}} \right]+\ldots \right) =
\sigma_{Born}
\left(1+\frac{3}{4}\frac{C_F\alpha_s(Q^2)}{\pi}+\ldots\right).
$$
The nature of the argument is purely perturbative. 
Can the Bloch-Nordsieck result hold beyond pQCD?  

Looking into this problem produced an extremely interesting result
that has laid a foundation for the development of perturbative\2
techniques aimed at analysing non-perturbative\1 effects.

V.~Braun, M.~Beneke and V.~Zakharov have demonstrated that the
real-virtual cancellation actually proceeds much deeper than was
originally expected. 

Let me briefly sketch the idea.  
\begin{itemize}
\item
 First one introduces an infrared cutoff (non-zero gluon mass $m$)
 into the calculation of the radiative correction.
\item
 Then, one studies the dependence of the answer on $m$.  A CIS
 quantity, by definition, remains finite in the limit $m\to0$.  This
 does not mean, however, that it is insensitive to the modification of
 gluon propagation. In fact, the $m$-dependence provides a handle for
 analyzing the {\em small transverse momenta}\/ inside Feynman
 integrals. It is this region of integration over parton momenta where
 the QCD coupling gets out of perturbative\1 control and the genuine
 non-perturbative physics comes onto the stage.
\item
 Infrared sensitivity of a given CIS observable is determined then by
 the first non-vanishing term which is {\em non-analytic}\/ in $m^2$ at
$m= 0$.
\end{itemize}
In the case of one-loop analysis of $\sigmatot$ that we are
discussing, one finds that in the sum of real and virtual
contributions not only the terms singular as $m \to 0$,
$$
\ln^2m^2\;,  \;\;\;\;\;\; \ln m^2
\,,
$$
cancel, as required by the Bloch-Nordsieck theorem, but that the
cancellation extends~\cite{BBZ,BB} also to the whole tower of {\em
finite}\/ terms
$$
m^2\ln^2 m^2\;, \;\;\;\; m^2\ln m^2\,, \;\;\;\; m^2\;, 
\;\;\;\; m^4\ln^2m^2\;, \;\;\;\; m^4\ln m^2 
\,. 
$$ 
In our case the first {\em non-analytic}\/ term appears at the level
of $m^6$:
$$ 
\frac{3}{4}\frac{C_F\alpha_s}{\pi}\left( 1 + 2\frac{m^6}{Q^6}\ln
\frac{m^2}{Q^2} +\cO{m^8}\right).
$$
It signals the presence of the non-perturbative $Q^{-6}$ correction to
$\sigmatot$, which is equivalent to that of the ITEP quark condensate
in~\eqref{eq:itep}. (The gluon condensate contribution emerges in the
next order in $\alpha_s$.)

A similar program can be carried out for other CIS quantities as well,
including intrinsically Minkowskian observables which address the
properties of the final state systems and, unlike the total cross
sections, do not have a Euclidean image.

The most spectacular non-perturbative\1 results were obtained for a
broad class of 
%%%%% so-called 
{\em jet shape variables}\/ (like thrust, $C$-parameter, broadenings,
and alike).  As has long been expected~\cite{hadro,DW,MW,AkZak}, these
variables possess relatively large $1/Q$ confinement correction
effects.

Employing the ``gluon mass'' as a large-distance trigger was
formalized by the so-called dispersive method~\cite{DMW}.
There it was also suggested to relate new non-perturbative\1
dimensional parameters with the momentum integrals of the effective
QCD coupling $\alpha_s$ in the infrared domain. Though it remains
unclear how such a coupling can be rigorously defined from the first
principles, the  {\em universality}\/ of the coupling 
%%%%% of the coupling 
makes this guess verifiable and therefore legitimate. All the
observables belonging to the same class $1/Q^{p}$ with respect to the
nature of the leading non-perturbative\1 behaviour, should be
described by the same parameter.

In particular, the extended family of jet shapes (including
energy-energy correlations~\cite{EEC}, out-of-plane transverse
momentum flows~\cite{kout} etc.) can be said to ``measure'' the first
moment of the perturbative\2 non-perturbative\1 coupling,
\begin{equation}
\alpha_0 \>\equiv\> \frac{1}{\mu_I}
%%%%% {2\,\mbox{\scriptsize GeV}} 
\int_0^{\mu_I}
%%%%% {2\,\mbox{\scriptsize GeV}}
 dk\,
\alpha_s(k^2), \qquad \mu_I= 2\,\mbox{GeV},
\end{equation}
where the choice of the ``infrared'' boundary value $\mu_I$ is a
matter of convention.

The interested reader will find a detailed discussion of the method,
of the guesses made and the problems faced, as well as the turbulent
history of its application, in review talks~\cite{rev}.  Here I will
only report the new spectacular results of the perturbative\2 study of
jet shape variables in DIS carried out recently by M.~Dasgupta and
G.~Salam~\cite{DISevsh}.

\section{INTERMEDIATE DISTANCES IN DIS}
%{Intermediate distances in DIS}
In Fig.~1 the results are shown of the two-parameter fits to the means
of jet shapes in $e^+e^-$ annihilation together with the fits to the
jet shape distributions in DIS~\cite{DISevsh}.
Consistency among the same-family variables is quite impressive.

\begin{figure}[h]
\begin{center}
%\vspace{3 cm}
 \epsfig{file=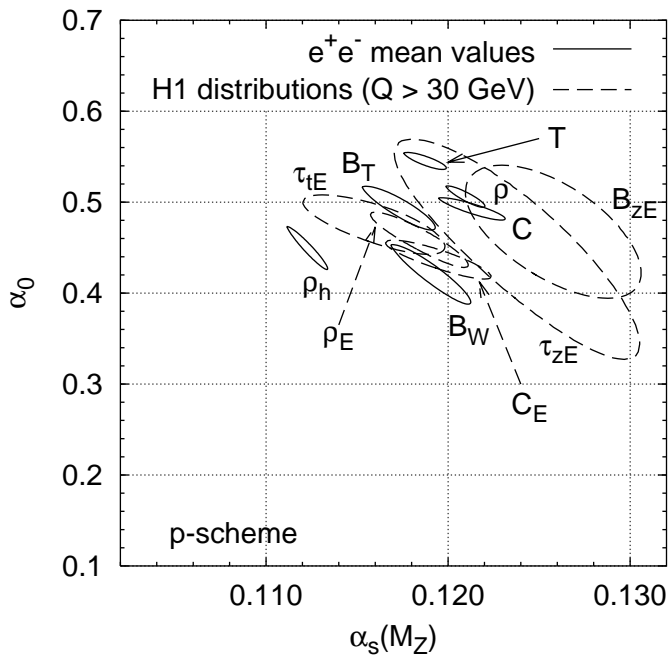}
\end{center}
\caption{$\alpha_s(M_Z^2)$ and $\alpha_0$ from jet shapes.
1--$\sigma$ contours for the means in $e^+e^-$ annihilation (solid)
and for the shape distributions in the current fragmentation jet in
DIS (dashed).}
\end{figure}

It is important to stress that prior to hunting for non-perturbative\1
effects, the state-of-the-art perturbative\1 predictions have to be
derived and implemented.  In the case of distributions this involves
resummation of logarithmically enhanced contributions in all order of
perturbation theory.  Having addressed this problem in the DIS
environment, Dasgupta and Salam have found a new set of log-enhanced
terms that has been previously overlooked in the literature. These
corrections (dubbed ``non-global'' by the founders) only affect the
observables that are based on a measurement restricted to a {\em
fraction}\/ of the total phase space available for gluon
radiation. For example, restricted to one hemisphere, or to any
limited angular region.
(In particular, among the observables that suffer from this newly
discovered effect is the Sterman-Weinberg jet cross section,--- the
first classical example of a CIS quantity.)

Being subleading (single logarithmic) in nature, these corrections
nevertheless modify the log-resummed perturbative predictions quite
significantly, as shown in Fig.~2.

\begin{figure}
\begin{center}
  \epsfig{file=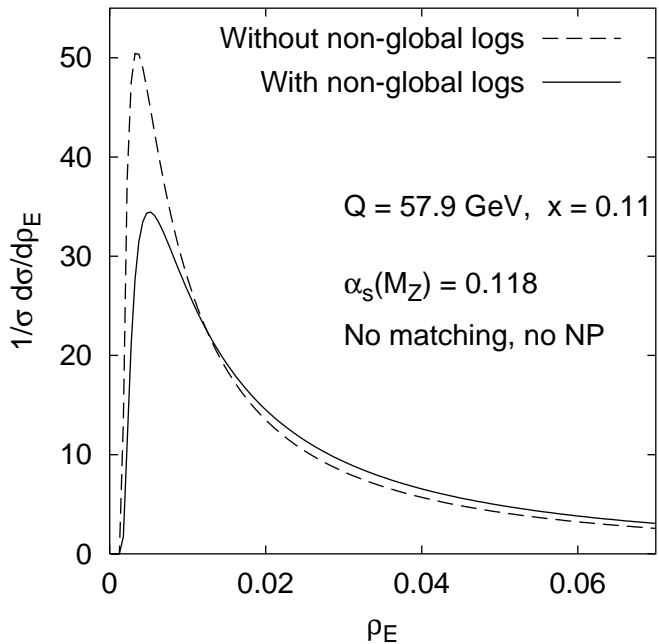}
\end{center}
\caption{Normalised invariant mass distribution with and without
non-global single-logarithmic corrections~\cite{NonGlobal}.}
\end{figure}

\section{CONCLUSIONS}
Deep inelastic scattering phenomena always were on the QCD forefront.
Exploring quark-gluon dynamics in the DIS environment becomes even
more important nowadays.
 
While we concentrated on {\em veryfying}\/ perturbative\1 QCD
predictions for multiple hadroproduction, DIS was handicapped as
lacking a manageable hadron-free initial state, as compared with the
``clean'' $e^+e^-$ annihilation.
Now that one aims at understanding an interface between hard and soft
physics, this is no longer a disadvantage, and DIS should take the
lead.

The main {\em advantage}\/ of DIS is that the {\em energy}\/ of
the process, $s$, is not kinematically equated to its hardness, $Q^2$,
as in the case of annihilation ($s=Q^2$).  Thus, in DIS one can study
intermediate and small hardness scales while staying away from the
difficult resonance region, without restricting the phase space for
multiparticle production.

On the other hand, the study of quasi-diffractive phenomena in
lepton-hadron scattering offers a variety of hardness handles ($Q^2$,
$t$, $J/\psi$ and $\Upsilon$ masses). Diffraction is interesting on
its own as a non-linear phenomenon closely linked to unitarity.
Moreover, it can be looked upon as a first step towards understanding
multi-gluon exchange, which is necessary for uncovering the
perturbative\2 non-perturbative\1 physics of lepton/hadron-nucleus and
heavy ion scattering.

\end{document}